# Site-selective magnetic moment collapse in compressed Fe$_5$O$_6$*

Qiao-Ying Qin(秦巧英)[1#], Ai-Qin Yang(杨爱芹)[1#], Xiang-Ru Tao(陶相如)[1], Liu-Xiang Yang(杨留响)[2], Hui-Yang Gou(缑慧阳)[2**], and Peng Zhang(张朋)[1**]

[1] MOE Key Laboratory for Nonequilibrium Synthesis and Modulation of Condensed Matter, School of Physics, Xi'an Jiaotong University, Xi'an, 710049, China
[2] Center for High Pressure Science and Technology Advanced Research, Beijing, 100094, China

*Supported by the National Natural Science Foundation of China under Grant No. 11604255, U1930401 and the Natural Science Basic Research Program of Shaanxi under Grant No. 2021JM-001.
# The two authors contribute equally to this work.
**Corresponding author. Email: zpantz@xjtu.edu.cn (P. Zhang)
                            huiyang.gou@hpstar.ac.cn(H.Y. Gou)

Iron oxide is one of the most important components in Earth's mantle. Recent discovery of the stable presence of Fe$_5$O$_6$ at Earth's mantle environment stimulates significant interests in the understanding of this new category of iron oxides. In this paper, we report the electronic structure and magnetic properties of Fe$_5$O$_6$ calculated by the density functional theory plus dynamic mean field theory (DFT+DMFT) approach. Our calculations indicate that Fe$_5$O$_6$ is a conductor at the ambient pressure with dominant Fe-3$d$ density of states at the Fermi level. The magnetic moments of iron atoms at three non-equivalent crystallographic sites in Fe$_5$O$_6$ collapse at significantly different rate under pressure. Such site-selective collapse of magnetic moments originates from the shifting of energy levels and the consequent charge transfer among the Fe-3$d$ orbits when Fe$_5$O$_6$ is being compressed. Our simulations suggest that there could be high conductivity and volume contraction in Fe$_5$O$_6$ at high pressure, which may induce anomalous features in seismic velocity, energy exchange, and mass distribution at the deep interior of Earth.

**PACS:** 75.40.Mg; 91.60.Gf; 91.60.Pn

    Iron oxide is one of the most abundant components in Earth's mantle, accounting for about 7.5% of the total mass of the Earth.[1,2] In Earth's mantle the temperature ranges from 500 K to 3000 K and the pressure ranges from 20 GPa to 140 GPa. The behavior of iron oxide at high-temperature and high-pressure is essential for understanding the evolution of Earth's interior.[3] At the ambient conditions, there are three known forms of iron oxides including wüstite (FeO), hematite (Fe$_2$O$_3$) and magnetite (Fe$_3$O$_4$).[4] Iron oxides have also been widely used in industry, [5-9] therefore attracted extensive interests in scientific communities.

    Recently, a series of new iron oxides were identified at high pressure and high temperature including FeO$_2$,[10,11] Fe$_4$O$_5$,[12] Fe$_5$O$_6$,[13] Fe$_5$O$_7$,[14] Fe$_7$O$_9$[14,15] and Fe$_{13}$O$_{19}$.[16] Among these iron oxides Fe$_5$O$_6$ is synthesized by iron and hematite fine powder at ratio ($Fe + 2Fe_2O_3 = Fe_5O_6$) in a diamond anvil chamber at pressure ranges from 10 GPa to 20 GPa and temperature at about 2000 K.[13] The experiment indicates that the crystal structure of Fe$_5$O$_6$ is orthorhombic (*Cmcm* space group) with lattice parameters of *a* = 5.319579 Bohr, *b* = 18.510432 Bohr, *c* = 28.367432 Bohr at 11.4 GPa.[13] There are three non-equivalent iron atoms and four unequal oxygen atoms in Fe$_5$O$_6$. As shown in Fig. 1, the iron atom at site-4*c* (Fe1) and the surrounding six oxygen atoms form a triangular prism, in contrast the other two iron atoms at site-8*f*1 (Fe2) and site-8*f*2 (Fe3) with their surrounding six oxygen atoms form octahedra respectively. Since Fe$_5$O$_6$ is phase stable under high temperature and high pressure, it is expected to be an important candidate material in

the interior of Earth along with other iron oxides. Experiment by Hikosaka *et al.* shows that $Fe_5O_6$ will decompose into $FeO + Fe_4O_5$ below 10 GPa and will decompose into $2FeO + Fe_3O_4$ above 38 GPa.[17] Recent experiment by Ovsyannikov *et al.* indicates the coexistence of the Verwey-type charge-ordering and the dimerization of iron atoms below 275 K,[18] where they find such charge ordering can be tuned by the Fe-Fe distance of the octahedral chain in $Fe_5O_6$. They further suggest that $Fe_5O_6$ could be used for memory devices or switches by controlling its charge ordering around the room temperature.

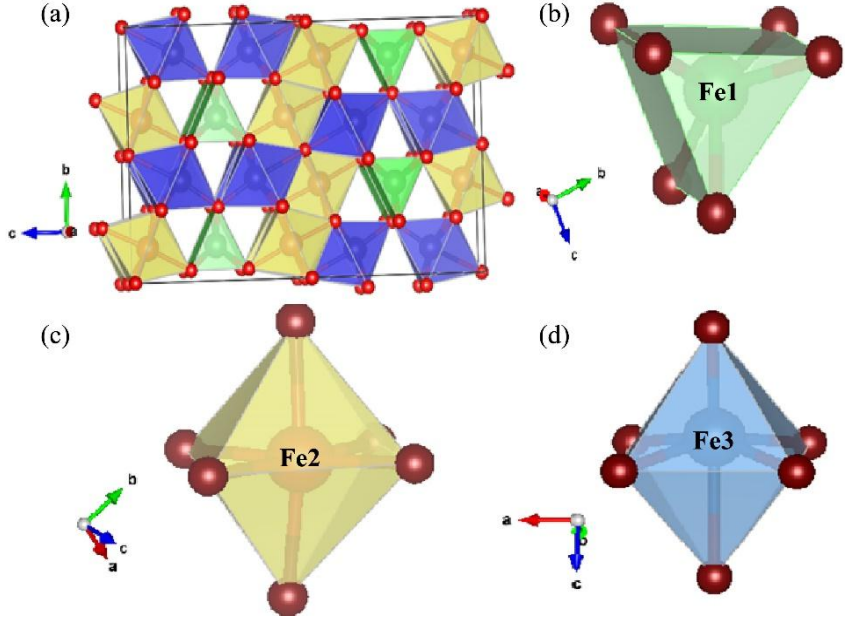

Fig. 1. (Color online) Crystal structure of $Fe_5O_6$ (a) the $FeO_6$ trigonal prism at site-4*c* (Fe1) in green color, (b) the $FeO_6$ octahedra at site-8*f1* (Fe2) in yellow color and (c) the $FeO_6$ octahedra at site-8*f2* (Fe3) in blue color.

The discovery of energetically stable $Fe_5O_6$ at high pressure opens up the possibility that it could be new candidate of Earth's mantle material. Unfortunately so far the experimental and the theoretical researches of $Fe_5O_6$ at the Earth's mantle conditions are very limited.[13,17,18] In this paper, we use the DFT+DMFT[19] method to calculate the electronic and the magnetic properties of paramagnetic $Fe_5O_6$ at high-pressure and high temperature. Our calculation proves that $Fe_5O_6$ is a metal with site-dependent magnetic moment collapse at high pressure. We further find that such site-dependent magnetic moment collapse originates from the energy level shifting and the consequent charge transfer among the five Fe-3*d* orbits in $Fe_5O_6$.

In DFT+DMFT calculations, we adopt the Wien2k package[20] in order to derive the charge density, the eigenvector and the eigenenergy of crystal in DFT level. The Wu-Cohen[21] exchange-correlation potential with $20 \times 20 \times 4$ k-points is used. The exact double counting method[22] and the hybrid expansion continuous-time quantum Monte Carlo (CTQMC) method [23-25] are used for solving the self-consistent DMFT equation. [26] In each DMFT iteration $10^7$ Monte Carlo updates are used and the self-energy is derived from the Dyson's equation. The converged self-energy from DMFT is used to update the new charge density and the new Kohn-Sham potential for the next DFT calculation. The $DFT + DMFT$ loops iterate until the fully convergence of the charge density, the self-energy and the hybridization functions *etc*. The maximum entropy method is used for analytic continuation of the self-energy from the imaginary frequency to the real frequency.[27]

We employ the constrained density functional theory method[28] to determine the screened Coulomb interaction U and the Hund's coupling J among the Fe-3*d* electrons. In orthorhombic $Fe_5O_6$ there are three non-equivalent iron atoms Fe1 (site-4*c*), Fe2 (site-8*f1*) and Fe3 (site-*8f2*). We derive the Coulomb U and the Hund's J at $V_0$=743.41 $bohr^3/Fe_5O_6$ and at V=698.34

bohr$^3$/Fe$_5$O$_6$ (at 11.4 GPa, also noted as $R = (1 - V/V_0)\% = 5.25\%$ representing the rate of volume compression). At V$_0$=743.41 bohr$^3$/Fe$_5$O$_6$ (U, J) = (5.15 eV, 1.03 eV) at Fe1, (5.08 eV, 0.97 eV) at Fe2, and (4.62 eV, 0.97 eV) at Fe3. At 11.4 GPa, (U, J) = (5.04 eV, 0.83 eV) at Fe1, (5.02 eV, 0.87 eV) at Fe2, and (4.45 eV, 0.95 eV) at Fe3. Since the value of the Coulomb interaction U and the Hund's coupling J is not sensitive to the sites of iron atoms and the volume of Fe$_5$O$_6$ per unit cell, throughout our DFT+DMFT calculations we choose the average (U, J) = (5.88 eV, 0.88 eV). DFT+U volume optimization with ($U - J = 5.0\ eV$) predicts V$_0$=743.41 bohr$^3$/Fe$_5$O$_6$ is the volume at the ambient pressure. In DFT+DMFT we further try (U, J) = (5.88 eV, 1.0 eV), (10.0 eV, 1.0 eV) and (10.0 eV, 0.88 eV) at the ambient conditions and find all the conclusions in this paper stay valid being independent of the specific values of U and J as presented above.

It is known that at ambient conditions iron oxides like FeO and Fe$_2$O$_3$ are insulators while in contrast Fe$_3$O$_4$ is metal. There is debate about the electronic conductibility of FeO$_2$.[29,30] Our DFT+DMFT calculations prove that Fe$_5$O$_6$ is a metal at the ambient pressure like Fe$_3$O$_4$. In Fig. 2, we present the total and the partial density of states (DOS) of Fe$_5$O$_6$. The total DOS of Fe$_5$O$_6$ have large quasi-particle peaks at the Fermi level in Fig. 2(a) at 743.41 bohr$^3$/Fe$_5$O$_6$ corresponding to the ambient pressure and in Fig. 2(b) at 408.88 bohr$^3$/Fe$_5$O$_6$ of 45% volume compression. This proves the metallic nature of Fe$_5$O$_6$ either at the ambient pressure or at high pressure. Also the height of quasi-particle peaks is relatively lower in Fig. 2(b) since the bandwidth of Fe$_5$O$_6$ is enlarged at high pressure. It is clearly shown in Fig. 2(c) and Fig. 2(d) that the total DOS of Fe$_5$O$_6$ at the Fermi level $E_F$ is dominated by the Fe-3$d$ bands, which suggests that the conductivity of Fe$_5$O$_6$ is mainly contributed by the Fe-3$d$ electrons.

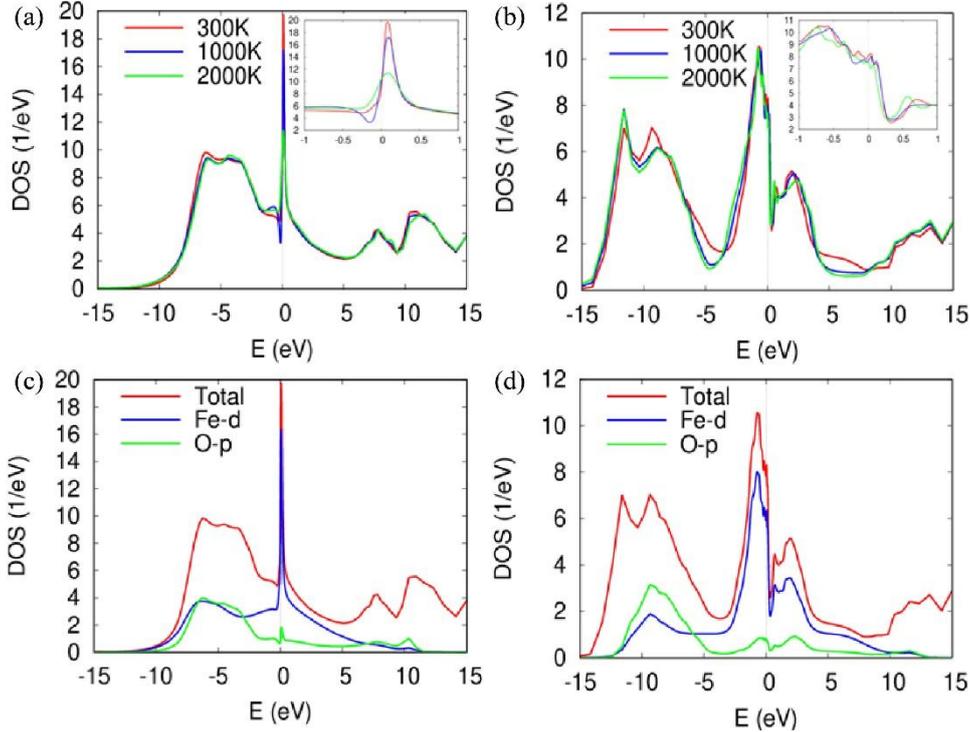

Fig. 2. The total density of states at 300 K, 1000 K, and 2000 K, (a) at the ambient pressure (V$_0$=743.41 bohr$^3$/Fe$_5$O$_6$) and (b) at R=45% volume compression (V=408.88 bohr$^3$/Fe$_5$O$_6$). The partial density of states of the Fe-3$d$ orbits and the O-p orbits at 300 K (c) at the ambient pressure and (d) at 45% volume compression.

There will be collapse of magnetic moment in iron oxides at high pressure.[30-36] Our DFT+DMFT calculations of the local magnetic moment $\sqrt{\langle m_z^2 \rangle}$ indicate that there is site-selective magnetic moment collapse in Fe$_5$O$_6$ under compression. As shown in Fig. 3, when Fe$_5$O$_6$ is compressed the local magnetic moment of Fe2 (8$f$1) decreases first, then the local magnetic moment of Fe3 (8*f*2), and the local magnetic moment of Fe1 (4*c*) is the last one to decrease. At the 45% volume compression endpoint, the magnetic moments of Fe2 (8*f*1) and Fe3 (8*f*2) saturate at about 0.8 $\mu_B/Fe$ and 1.1 $\mu_B/Fe$ respectively, in contrast the magnetic moment of Fe1 (4*c*) ends at about 1.4 $\mu_B/Fe$. The dependence of the local magnetic moment on temperature and volume compression is summarized in the three diagrams in Fig. 4. It is clear that the magnetic moments of the three non-equivalent iron atoms always decrease as being compressed at any temperature from 300 K to 2000 K. Either at low temperature or at high temperature, the magnetic moment of Fe atoms collapses first at 8*f*1 position, and then 8*f*2 position, the last is at 4c position. In Fig. 4(a), 4(b) and 4(c), the green region that represents the magnetic moment of intermediate size expands as the temperature is increased, which indicates that decrement of the magnetic moment is slower at high temperature.

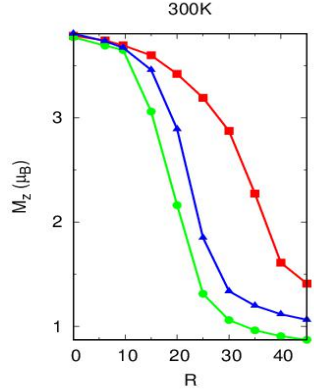

Fig. 3. The local magnetic moment of three non-equivalent iron atoms in $Fe_5O_6$ as a function of volume compression rate at 300 K. The red, the green, and the blue lines represent the local magnetic moments of Fe1 (site-4*c*), Fe2 (site-8*f*1), and Fe3 (site-8*f*2) respectively.

In transition metal oxides the collapse of magnetic moment is often accompanied by the insulator-metal transition and the collapse of volume.[32-36] Thus the insulator-metal transition is presumed to be related to the collapse of magnetic moment. However in $Fe_5O_6$ we show the site-selective high spin-low spin transition of three iron atoms can be independent of the insulator-metal transition since $Fe_5O_6$ is always metallic from the ambient pressure to high pressure. Similar behavior has also been found in $FeO_2$.[30] It rises up the question that whether the metallization of Mott insulator is the consequence of the pressure driven high spin-low spin transition.

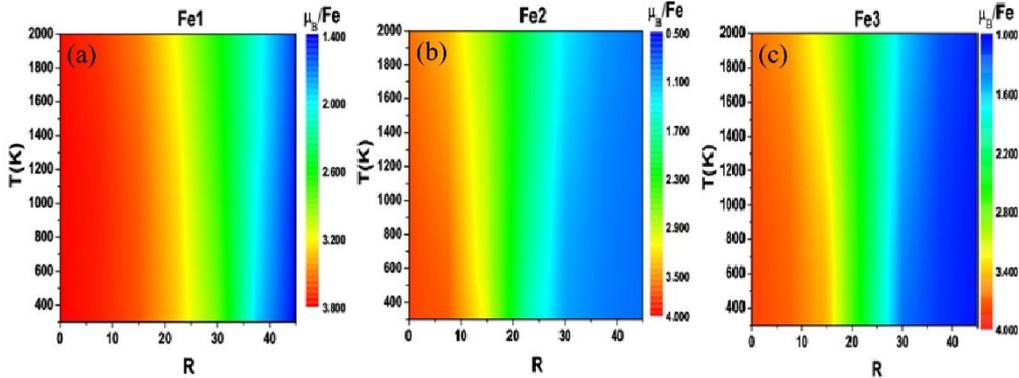

Fig.4. The dependence of local magnetic moments on temperature and volume compression rate at the three non-equivalent positions of $Fe_5O_6$ (a) Fe1 (site-4*c*), (b) Fe2 (site-8*f*1), and (c) Fe3 (site-8*f*2).

It has been proved that the magnetic moment collapse in transition metal oxides such as FeO and MnO originates from the transfer of electrons among the *d*-orbits of the transition metals.[32-34,36] As shown in Fig. 5, we present the Fe-3d electron occupancies at Fe1 (4*c*), Fe2 (8*f*1) and Fe3 (8$f$2) of Fe$_5$O$_6$ at 300 K. When volume compression rate increases, at all three crystallographic non-equivalent sites the electron occupancies of the $z^2$, $x^2-y^2$ and $xz$ orbits increase, while the electron occupancies of the $yz$ and $xy$ orbits decrease. It indicates the charge transfer from the yz and xy orbits to the $z^2$, $x^2-y^2$ and $xz$ orbits upon the volume compression. However, for Fe1 (4$c$), the charge transfer from the $yz$ and $xy$ orbits to the $z^2$ and $xz$ orbits start at relatively larger volume compression rate R than that of Fe2 (8$f$1) and Fe3 (8$f$2), and the electron occupancy of the $x^2-y^2$ is almost constant until R is above 40%. This indicates that the electrons at Fe1 (4$c$) have stronger correlations than the electrons at Fe2 (8$f$1) and Fe3 (8$f$2). Such electron transfer among the Fe-3$d$ orbits is temperature dependent in that at higher temperature the charge will transfer at a lower rate.

The relationship between the local magnetic moment collapses of iron atoms and the charge transfer among the Fe-3$d$ orbits could be understood at the atomic limit. At the low pressure end all five Fe-3$d$ orbits have almost the same number of electrons and the magnetic moments is maximized at around 4.0 $\mu_B/Fe$ due to the Hund's rule. At the high pressure limit when the $yz$ and the $xy$ orbits lost all their electrons, then the six electrons on the $z^2$, $x^2-y^2$ and $xz$ orbits must form singlet pairs because of the Pauli's exclusion principle which leads to zero total magnetic moment.

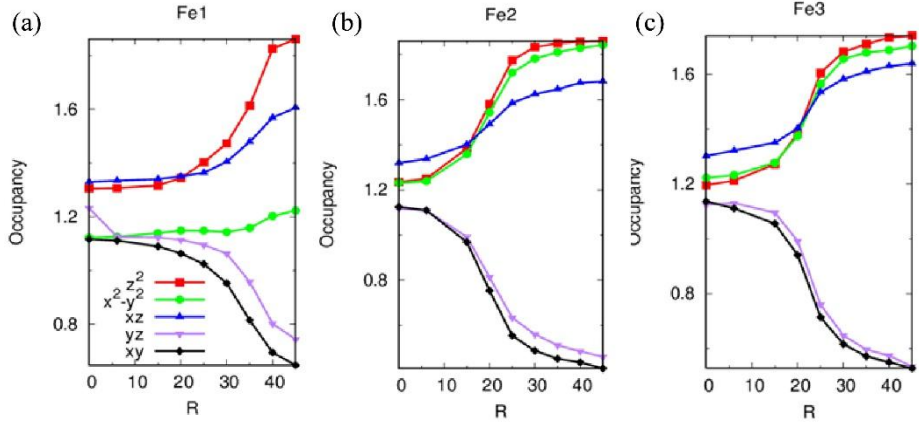

Fig.5. The electron occupancy number of the Fe-3$d$ orbits of atom (a) Fe1 (site-4$c$), (b) Fe2 (site-8$f$1) and (c) Fe3 (site-8$f$2) as a function of volume compression at 300 K.

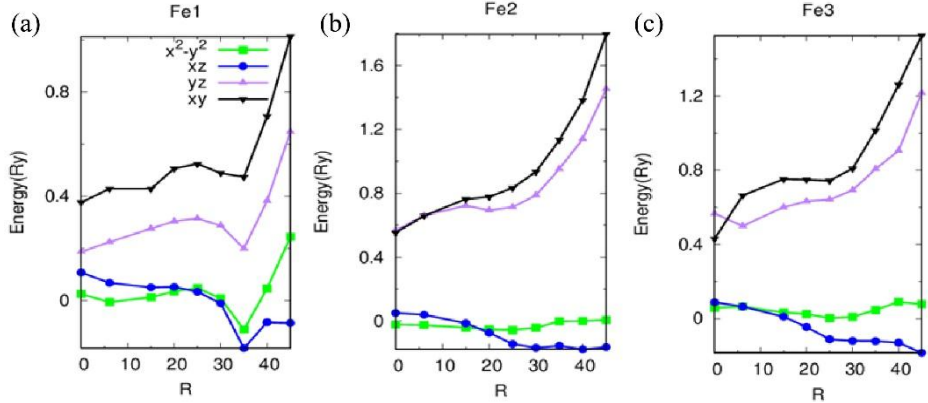

Fig.6. The energy differences of the Fe-3$d$ orbits of atom (a) Fe1 (site-4$c$), (b) Fe2 (site-8$f$1), and (c) Fe3 (site-8$f$2) as a function of volume compression rate at 300 K.

The transfer of electrons among the five Fe-3$d$ orbits originates from the shift of energy levels of the Fe-3$d$ orbits under compression. The energy difference of the $x^2-y^2$, $xz$, $yz$ and $xy$ orbits relative to the $z^2$ orbit under compression is shown in Fig. 6. In Fig. 6 (b) and Fig. 6 (c), upon compression energies of the $yz$ and $xy$ orbits at Fe2 (8$f$1) and Fe3 (8$f$2) sites moves upward, however energies of the $x^2-y^2$ orbits stay almost invariant and energy of $xz$ even decrease. Then naturally electrons on the $yz$ and $xy$ orbits transfer to the $z^2$, $x^2-y^2$, and $xz$ orbits in order to minimize the total energy in $Fe_5O_6$. In contrast in Fig. 6 (a) the orbital-energy differences of the $yz$ and $xy$ orbits of Fe1 (4$c$) increase much slower under compression, which leads to the slower magnetic moment collapse as observed in Fig. 3.

In summary, we examine the electromagnetic properties of the recently discovered iron oxides $Fe_5O_6$ at the high pressure high temperature conditions of Earth's interior employing the DFT+DMFT method. Our calculations prove that the paramagnetic $Fe_5O_6$ is metallic either at the ambient pressure or at high pressure. The electronic conduction in $Fe_5O_6$ is dominated by the Fe-3$d$ electrons. The local magnetic moments of three crystallographic non-equivalent iron atoms in $Fe_5O_6$ collapse at different rate as being compressed, which leads to site-dependent magnetic moment collapse. We further find that the site-dependent magnetic moment collapse in $Fe_5O_6$ originates from the energy level shifting and the consequent charge transfer among the five Fe-3$d$ bands under volume compression. Since $Fe_5O_6$ is a candidate component of Earth's mantle beside the insulator FeO, the metallic nature of $Fe_5O_6$ suggests the conductivity of Earth's mantle could be higher than previously expected. The electronic conductivity and the related thermal conductivity of Earth's mantle may strongly affect the dynamics in Earth's deep interior. For example they could determine the energy exchange rate and the mass distribution of Earth's low mantle-core boundary, which may induce anomalous features in seismic velocity. Thus further investigations on $Fe_5O_6$ and other new iron oxides are called for the precise value of the electronic conductivity and the thermal conductivity at Earth's interior environment.

*Acknowledgments*. We thank the High Performance Computing platform of Xi'An Jiaotong University for providing the allocation of CPU time.